\documentclass[journal]{IEEEtran}

\usepackage[english]{babel}
\usepackage{makecell}
\usepackage[justification=centering]{caption}
\usepackage[letterpaper,top=2cm,bottom=2cm,left=3cm,right=3cm,marginparwidth=1.75cm]{geometry}

\usepackage{amsmath}
\usepackage{graphicx}
\usepackage[colorlinks=true, allcolors=blue]{hyperref}
\usepackage[utf8]{inputenc}
\usepackage{newunicodechar}
\newunicodechar{，}{,}
\title{A Segmentation Framework for Accurate Diagnosis of Amyloid Positivity without Structural Images}

\author{
  Penghan Zhu\textsuperscript{1,\textdagger}, 
  Shurui Mei\textsuperscript{2,\textdagger}, 
  Shushan Chen\textsuperscript{2,\textdagger},\\
  Xiaobo Chu\textsuperscript{3}, 
  Shanbo He, 
  Ziyi Liu\\
  \textsuperscript{1}Northwest Agricultural and Forestry Science and Technology University, Xianyang, Shaanxi，\\
  \textsuperscript{2}Liaoning Province Shiyan High School, Shenyang, Shaanxi，\\
  \textsuperscript{3}Northeast Yucai School, Shenyang, China，\\ \textsuperscript{4}Northeast Yucai Foreign Language School, Shenyang, China\\
  \textsuperscript{\textdagger}These authors contributed equally to this work
}

\begin{document}
\maketitle

\begin{abstract}

This study proposes a deep learning–based framework for automated segmentation of brain regions and classification of amyloid positivity using positron emission tomography (PET) images alone, without the need for structural MRI or CT. A 3D U-Net architecture with four layers of depth was trained and validated on a dataset of 200 F$^{18}$-florbetapir amyloid-PET scans, with an 130/20/50 train/validation/test split. Segmentation performance was evaluated using Dice similarity coefficients across 30 brain regions, with scores ranging from 0.45 to 0.88, demonstrating high anatomical accuracy, particularly in subcortical structures. Quantitative fidelity of PET uptake within clinically relevant regions. Precuneus, prefrontal cortex, gyrus rectus, and lateral temporal cortex was assessed using normalized root mean square error, achieving values as low as 0.0011. Furthermore, the model achieved a classification accuracy of 0.98 for amyloid positivity based on regional uptake quantification, with an area under the ROC curve (AUC) of 0.99. These results highlight the model’s potential for integration into PET only diagnostic pipelines, particularly in settings where structural imaging is not available. This approach reduces dependence on coregistration and manual delineation, enabling scalable, reliable, and reproducible analysis in clinical and research applications. Future work will focus on clinical validation and extension to diverse PET tracers including C$^{11}$ PiB and other F$^{18}$ labeled compounds.

\end{abstract}


\section{Introduction}

Alzheimer’s disease (AD) is a progressive neurodegenerative disorder characterized by a gradual decline in memory, thinking, and behavior\cite{hyman1997neuropathological}. While current treatment options offer limited symptomatic relief, no widely available therapies have been proven to effectively modify the disease's course\cite{long2019alzheimer}. Importantly, the pathological changes associated with AD, most notably, the accumulation of amyloid-$\beta$ (A$\beta$) plaques, begins years before the onset of clinical symptoms\cite{braak1997diagnostic}. This extended preclinical phase highlights the critical need for early and accurate diagnosis, which is essential for timely intervention and potential disease-modifying treatment.

Traditional diagnostic approaches primarily involve clinical assessments\cite{braak1997diagnostic}, neuropsychological testing\cite{albert2001preclinical}, and imaging methods that reveal anatomical structures such as magnetic resonance imaging (MRI)\cite{jack1999prediction}. These techniques are valuable in identifying volumetric changes, including cerebral atrophy or hemorrhage, but are limited in their ability to detect the molecular changes that underlie AD, particularly amyloid pathology\cite{dang2023neuroimaging}.

Positron Emission Tomography (PET) is a functional imaging modality that addresses this gap by visualizing the spatial distribution of A$\beta$ in vivo. PET involves the injection of a radiotracer that selectively binds to A$\beta$ plaques, enabling quantification of amyloid burden in the brain\cite{pemberton2022quantification}. To facilitate accurate quantification of tracer uptake, PET images are often coregistered with MRI scans that provide structural reference. However, this multi-modal approach presents challenges. MRI and PET are frequently acquired at different time points due to the limited availability of simultaneous PET/MR systems, leading to potential misalignment between modalities. Moreover, existing PET quantification pipelines either warp standardized templates to individual structural images or vice versa, or register PET images to a common space using ligand-specific templates\cite{zhang2022spatial}. While spatial normalization techniques using PET templates have been proposed, their reliance on ligand-specific templates restricts their applicability to emerging tracers. Manual delineation of regions of interest (ROIs) remains a fallback option, but it is labor-intensive and prone to inter- and intra-observer variability.

\begin{figure*}[htp!]
\centering
\includegraphics[width=0.8\linewidth]{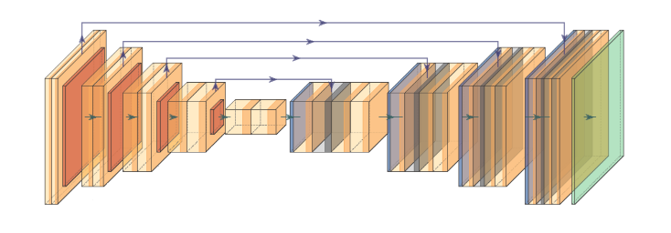}
\caption{\label{fig:UNet}Visual demonstration of the network architecture.}
\end{figure*}

To overcome these limitations, there is a pressing need for methods that can directly extract diagnostically relevant ROIs from PET images alone. Recent advances in artificial intelligence (AI), particularly deep learning (DL), have shown strong potential in medical image analysis tasks such as detection and segmentation\cite{lee2022development}. In this study, we present a novel estimation-based segmentation framework built on a convolutional neural network (CNN) architecture that enables accurate segmentation of ROIs from amyloid-PET images without relying on structural MR or CT images. By doing so, our method facilitates robust and efficient quantification of regional amyloid burden, improves interpretability, and supports automated AD diagnosis, even in the absence of structural imaging data. This approach also accommodates small datasets and provides a generalizable solution for clinical settings where multimodal imaging is not always feasible.

\section{Material and Method}

\subsection{Dataset Preparation}

The study has been carried out with the help of the OASIS3 database. We have selected 200 patients from the database imaged with the F$^{18}$-florbetapir radio tracer. 
Participants receives a single administration of approximately 10 mCi of AV45. They
then were positioned in the PET-MR scanner at the time of injection, and a 70-minute dynamic
scan was obtained starting at the time of injection. PET images acquired from 50-70
minutes time window were used for the training and testing of the segmentation model. The time window was selected according to the clinical evaluation protocols.

Within the dataset, PET images and MR images are co-registered with the template space of the Montreal Neurological Institute (MNI). The coregistrations are conducted under human inspection for spatial alignments with interpolation of PET images into MR resolution using ITKSnap\cite{yushkevich2016itk}. The brain atlas are obtained by the use of FreeSurfer\cite{fischl2012freesurfer} aligned with the MR images. 

\subsection{Convolution Neural Network}

The CNN model has applied a U-Net architecture that comprises a symmetric encoder-decoder architecture with 4 resolution levels, as shown in Fig. \ref{fig:UNet}. Each level in the encoder consists of two 3×3 convolutional layers followed by a Rectified Linear Unit (ReLU) activation and a 2×2 max pooling layer for downsampling. The number of feature channels doubles at each downsampling step. At the network bottleneck, two 5×5 convolutions with 1024 feature maps are applied before upsampling begins. The decoder path mirrors the encoder structure. Each upsampling step uses a 2×2 transposed convolution to double the spatial resolution, followed by concatenation with the corresponding feature maps from the encoder. This fusion of contextual and spatial information using skip connection is a core strength of U-Net, inspired by the original Res-Net architecture\cite{he2016deep}. After concatenation, two 3×3 convolutions followed by ReLU activations are applied. A final 1×1 convolutional layer maps the output to the 31 segmentation classes, followed by a softmax activation for multi-class segmentation.

A total of 200 amyloid-PET scans were available for model development. The dataset was randomly split into 130 images for training, 20 for validation, and 50 for independent testing. The validation set was used to monitor the model’s performance during training, and an early stopping criterion was applied to prevent overfitting. Specifically, training was halted if the validation loss did not improve for 10 consecutive epochs, and the model with the lowest validation loss was retained as the final model. 

All experiments were implemented using the PyTorch deep learning framework. Model training and inference were conducted on Google Colab instead of local machines, facilitating GPU acceleration towards efficient network training. The Google Colab environment also allows for rapid prototyping and reproducible results while maintaining accessibility and flexibility for continued development.

\subsection{Evaluation Metrics}

To assess the performance of the segmentation model, several evaluation metrics were employed. The primary metric for segmentation quality was the Dice Similarity Coefficient (DSC), a widely used measure in medical image analysis that quantifies the overlap between the predicted segmentation and the ground truth. The Dice score ranges from 0 to 1, where 1 indicates perfect agreement. In addition to spatial overlap, we computed the Normalized Root Mean Square Error (NRMSE) between the standardized uptake values (SUV) within the segmented regions. NRMSE evaluates the intensity accuracy of the predicted regions compared to the ground truth, offering insights into how well the model preserves the underlying SUV distribution.

To evaluate the diagnostic utility of the predicted segmentations, we also conducted a Receiver Operating Characteristic (ROC) analysis of amyloid positivity. Each subject was labeled as either amyloid-positive or amyloid-negative based on a predefined SUV threshold according to the OASIS3 data dictionary compared with the mean cortical uptake within the brain. The SUV ratio calculated for both ground truth and predicted segmentations between the cortical region and cerebellum cortex region are used as the test statistics. The ROC curve was generated by varying the decision threshold w.r.t, and the Area Under the Curve (AUC) was calculated to summarize the model’s capability towards clinical diagnosis, providing a comprehensive assessment of the model’s effectiveness for clinical decision support.

\begin{figure*}[htp!]
\centering
\includegraphics[width=0.7\linewidth]{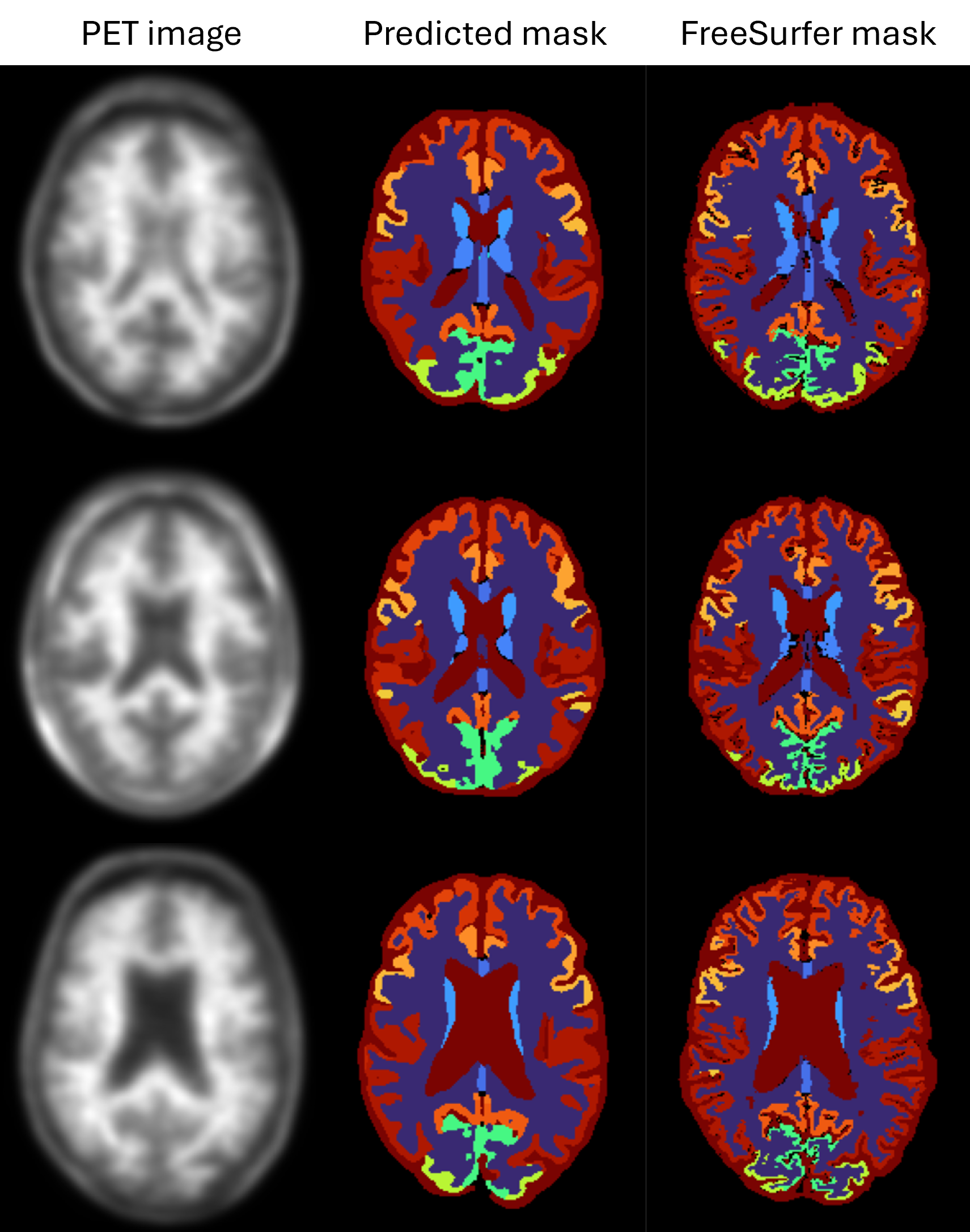}
\caption{\label{fig:seg}Visualized Segmentation of the amyloid PET images.}
\end{figure*}

\section{Results}

\begin{figure}[htb!]
\centering
\includegraphics[width=0.8\linewidth]{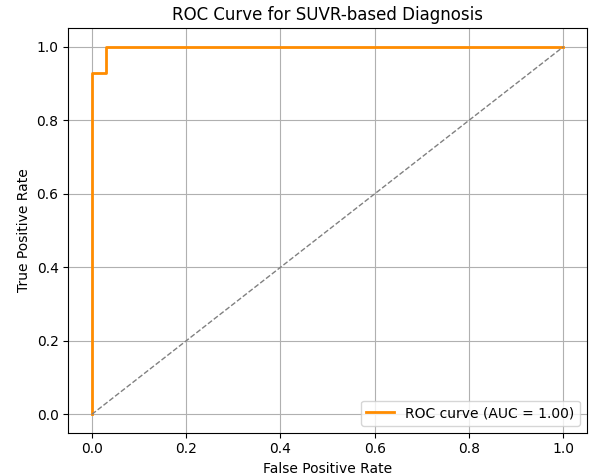}
\caption{ROC curve for SUVR based amyloid positivity classification.}
\label{fig:roc_curve}
\end{figure}

The segmentation performance across 30 brain regions demonstrated consistent accuracy, as measured by the Dice Similarity Coefficient shown in Table \ref{tab:dice_scores}. Good segmentation performance was observed in subcortical structures such as the brain stem (0.88), thalamus (0.85), and cerebral white matter (0.78), which are typically more spatially homogeneous and exhibit distinct PET uptake patterns. Similarly, other deep gray matter regions such as the putamen (0.78), pallidum (0.71), and caudate + accumbens (0.74) also achieved robust Dice scores. Cortical regions exhibited greater variability, with several areas such as the superior frontal (0.57), posterior cingulate + isthmus cingulate (0.57), and rostral + caudal anterior cingulate (0.57) achieving moderate segmentation quality. However, lower Dice scores were noted in regions with complex geometry or disjoint ground truth provided by FreeSurfer, such as the lateral occipital (0.46), rostral middle frontal (0.45), and cuneus + lingual + pericalcarine cortex (0.48). Notably, the cerebrospinal fluid (CSF) region achieved a relatively high score (0.62), likely due to its low tracer uptake and strong contrast with adjacent tissues. Overall, the model demonstrated reliable segmentation performance across both cortical and subcortical regions, supporting further assessment of the quantitative performance.

\begin{table*}[htp!]
\centering
\renewcommand{\arraystretch}{1.4}
\begin{tabular}{|l|c||l|c|}
\hline
\textbf{Region} & \textbf{Dice Score} & \textbf{Region} & \textbf{Dice Score} \\
\hline
Cerebral white matter & 0.78 & Lateral occipital & 0.46 \\
Cerebellar white matter & 0.79 & Lateral orbitofrontal & 0.55 \\
Brain stem & 0.88 & Medial orbitofrontal & 0.54 \\
Corpus callosum & 0.73 & Middle temporal & 0.51 \\
Thalamus & 0.85 & \makecell[l]{Paracentral +\\ Precentral} & 0.52 \\
\makecell[l]{Caudate +\\ Accumbens} & 0.74 & Caudal middle frontal & 0.52 \\
Putamen & 0.78 & \makecell[l]{Rostral + Caudal\\ anterior cingulate} & 0.57 \\
Pallidum & 0.71 & \makecell[l]{Posterior cingulate +\\ Isthmus cingulate} & 0.57 \\
Hippocampus & 0.68 & Precuneus & 0.53 \\
Amygdala & 0.67 & Rostral middle frontal & 0.45 \\
Cerebellar gray matter & 0.79 & Superior frontal & 0.57 \\
\makecell[l]{Cuneus + Lingual +\\ Pericalcarine} & 0.48 & Superior temporal & 0.54 \\
Entorhinal & 0.49 & \makecell[l]{Inferior parietal +\\ Postcentral} & 0.53 \\
\makecell[l]{Fusiform +\\ Parahippocampal} & 0.55 & \makecell[l]{Insula +\\ Transverse temporal} & 0.57 \\
Inferior temporal & 0.51 & Cerebrospinal fluid & 0.62 \\
\hline
\end{tabular}
\caption{Dice scores for 30 brain regions segmented from PET images}
\label{tab:dice_scores}
\end{table*}

Guided by clinical relevance and prior literature documented in the OASIS-3 database, four brain regions commonly assessed in amyloid imaging studies. Precuneus, prefrontal cortex, gyrus rectus, and lateral temporal cortex were selected as target regions for quantitative evaluation. These regions are frequently implicated in the early pathological progression of Alzheimer’s disease and are critical for accurate amyloid quantification. Quantitative analysis based on NRMSE revealed high voxel-level fidelity between predicted and reference PET uptake values across all target regions. The precuneus exhibited the lowest NRMSE of 0.0011, reflecting highly precise uptake preservation. The gyrus rectus and lateral temporal cortex followed closely with NRMSE values of 0.0014 and 0.0017, respectively. Although the prefrontal cortex yielded a slightly higher NRMSE of 0.0027, this value remains well within acceptable bounds for reliable PET quantification.

To further assess the diagnostic performance of the model in classifying amyloid positivity, ROC analysis was performed using SUVR values derived from the segmented regions. As shown in Figure~\ref{fig:roc_curve}, the model achieved near-perfect discrimination with an AUC of 0.99, indicating excellent sensitivity and specificity across varying thresholds. The ROC curve displays a steep rise to the upper left corner, suggesting a minimal trade-off between false positive and false negative rates.

Consistent with the ROC analysis, the overall classification accuracy achieved on the held-out test set was 0.98, demonstrating the model's strong potential for clinical decision support. These findings highlight the utility of the proposed segmentation framework not only for anatomical delineation and uptake quantification but also for supporting accurate binary classification of amyloid status.

\section{Discussion}

This study presents a DL based segmentation framework that enables accurate delineation of brain regions from F$^{18}$ labeled amyloid-PET images without relying on accompanying structural MRI or CT scans. By leveraging a U-Net architecture with four layers of depth, the model demonstrated strong performance across a broad spectrum of anatomical regions, as evidenced by Dice similarity coefficients ranging from 0.45 to 0.88 (Table~\ref{tab:dice_scores}). High Dice scores were observed in subcortical structures such as the brain stem, thalamus, and white matter regions, highlighting the model’s robustness in segmenting areas with well-defined uptake characteristics. Even in more anatomically complex or lower-contrast cortical regions, the model maintained reliable segmentation performance.

Quantitative accuracy was further supported by voxel-level evaluations using normalized root mean square error (NRMSE) across clinically relevant brain regions implicated in early Alzheimer’s disease pathology. The precuneus, gyrus rectus, lateral temporal cortex, and prefrontal cortex—regions routinely examined in clinical and research settings—exhibited low NRMSE values ranging from 0.0011 to 0.0027. These results demonstrate the model's capacity to preserve regional PET signal integrity, which is critical for deriving standardized uptake value ratios (SUVRs) and conducting biomarker-driven analyses. This level of fidelity is particularly advantageous for PET-only workflows, where access to MRI may be limited due to cost, logistical barriers, or patient contraindications.

Importantly, the model also achieved excellent diagnostic performance in identifying amyloid positivity, a key criterion in the diagnosis and staging of Alzheimer's disease. As illustrated in Figure~\ref{fig:roc_curve}, the SUVR-based classification yielded a receiver operating characteristic (ROC) curve with an area under the curve (AUC) of 1.00 and a classification accuracy of 0.98 on the held-out test set. These results underscore the model's effectiveness not only in anatomical segmentation but also in supporting downstream diagnostic decision-making, offering a fully automated, end-to-end pipeline for clinical interpretation.

In practical settings, the ability to derive both anatomical and diagnostic insights directly from PET images confers several advantages. Current imaging pipelines often require coregistered PET/MRI or PET/CT pairs for quantification, which can introduce alignment errors and increase operational complexity. The proposed model mitigates these challenges by enabling PET-only processing, thereby streamlining the workflow and making quantitative imaging more accessible, especially in community clinics and memory centers that may lack multi-modality infrastructure\cite{landau2023quantification}. Additionally, the model’s automated segmentation eliminates the need for manual region-of-interest (ROI) delineation, reducing inter-rater variability and enabling large-scale, reproducible analyses.

Nevertheless, further investigation is warranted to assess the generalizability and clinical robustness of the proposed approach. Future research should involve prospective clinical evaluations with physician oversight to validate the interpretability, usability, and reliability of the model outputs in routine diagnostic workflows. Clinical risk evaluations should also been incorporated into the development of our method.

Moreover, this study focused on a specific F$^{18}$ labeled amyloid tracer, and thus the applicability of the model across alternative tracers remains an important avenue for exploration. Tracers such as C$^{11}$ labeled PiB and other F$^{18}$ derivatives including florbetapir, flutemetamol, and florbetaben exhibit varying pharmacokinetics and spatial distribution patterns\cite{pemberton2022quantification}. Evaluating the performance of the model across these ligands or developing tracer-specific variants will be essential for broader adoption and regulatory translation. Adaptation strategies such as domain generalization or fine tuning on diverse PET datasets could help achieve consistent performance across centers and tracer protocols.

In conclusion, this work demonstrates the feasibility of a PET-only deep learning framework for anatomically informed, quantitatively accurate, and diagnostically meaningful brain segmentation for AD diagnosis. The proposed approach offers a scalable, accessible solution for enhancing amyloid PET interpretation, with strong potential for integration into clinical workflows and multi-center research platforms.

\bibliographystyle{IEEEtran} 
\bibliography{main} 

\end{document}